%% file: ms.tex
\documentclass[12pt,twocolumn,tighten]{aastex62}
\usepackage{amsmath,amstext,amssymb}
\usepackage[T1]{fontenc}
\usepackage{apjfonts}
\usepackage[figure,figure*]{hypcap}
\usepackage{graphics,graphicx}
\usepackage{hyperref}
\usepackage{natbib}

\received{February 28, 2020}
\revised{March 27, 2020}
\accepted{March 30, 2020}
\submitjournal{The Astrophysical Journal Letters}
\shorttitle{WASP-4 is accelerating toward the Earth}

\begin{document}

\defcitealias{bouma_wasp4b_2019}{B19}

\title{WASP-4 is Accelerating Toward the Earth}

\correspondingauthor{L. G. Bouma}
\email{luke@astro.princeton.edu}

%
%
\author[0000-0002-0514-5538]{L. G. Bouma}
\affiliation{ Department of Astrophysical Sciences, Princeton
University, 4 Ivy Lane, Princeton, NJ 08540, USA}
\author[0000-0002-4265-047X]{J. N. Winn}
\affiliation{ Department of Astrophysical Sciences, Princeton
University, 4 Ivy Lane, Princeton, NJ 08540, USA}

%
%
\author[0000-0001-8638-0320]{A. W. Howard}
\affiliation{Cahill Center for Astrophysics, California Institute of
Technology, Pasadena, CA 91125, USA}
\author[0000-0002-2532-2853]{S. B. Howell}
\affiliation{NASA Ames Research Center, Moffett Field, CA 94035, USA}
\author[0000-0002-0531-1073]{H. Isaacson}
\affiliation{Astronomy Department, University of California, Berkeley,
CA 94720, USA}
\author{H. Knutson}
\affiliation{Division of Geological and Planetary Sciences, California
Institute of Technology, Pasadena, CA 91125, USA}
\author[0000-0001-7233-7508]{R. A. Matson}
\affiliation{U.S. Naval Observatory, Washington, DC 20392, USA}

\begin{abstract}
  The orbital period of the hot Jupiter WASP-4b appears to be
  decreasing at a rate of $-8.64 \pm 1.26~{\rm msec}\,{\rm yr}^{-1}$,
  based on transit-timing measurements spanning 12 years.  Proposed
  explanations for the period change include tidal orbital decay,
  apsidal precession, and acceleration of the system along the line of
  sight.  To investigate further, we performed new radial velocity
  measurements and speckle imaging of WASP-4.  The radial-velocity
  data show that the system is accelerating towards the Sun at a rate
  of $-0.0422\pm 0.0028\,{\rm m}\,{\rm s}^{-1}\,{\rm day}^{-1}$.  The
  associated Doppler effect should cause the apparent period to shrink
  at a rate of $-5.94 \pm 0.39~{\rm msec}\,{\rm yr}^{-1}$, comparable
  to the observed rate.  Thus, the observed change in the transit
  period is mostly or entirely produced by the line-of-sight
  acceleration of the system.  This acceleration is probably caused by
  a wide-orbiting companion of mass 10-300$\,M_{\rm Jup}$ and orbital
  distance 10-100$\,$AU, based on the magnitude of the radial-velocity
  trend and the non-detection of any companion in the speckle images.
  We expect that the orbital periods of 1 out of 3 hot Jupiters will
  change at rates similar to WASP-4b, based on the hot-Jupiter
  companion statistics of \citet{knutson_friends_2014}.
  Continued radial velocity monitoring of hot Jupiters is therefore
  essential to distinguish the effects of tidal orbital decay or
  apsidal precession from line-of-sight acceleration.
\end{abstract}

\keywords{Exoplanet tides (497), Exoplanet dynamics (490), Radial
velocity (1332), Transit timing variation method (1710)}


\section{Introduction}

The orbits of most hot Jupiters are formally unstable to tidal decay
\citep{counselman_outcomes_1973,hut_stability_1980,rasio_tidal_1996,levrard_falling_2009,matsumura_tidal_2010}.
It is not clear, though, whether the timescale for tidal orbital decay
is shorter or longer than the timescale for main-sequence stellar
evolution.  This answer to this question depends on the uncertain rate
at which friction inside the star damps the tidal oscillations, which
is ultimately what causes the orbit to shrink (as reviewed by
\citealt{Mazeh2008} and \citealt{ogilvie_tidal_2014}).  Population
studies of hot Jupiters --- based on ages, rotation rates, orbital
distances, and Galactic kinematics --- have led to differing
conclusions, with estimated lifetimes ranging from less than a
gigayear to much longer than main-sequence lifetimes (see, {\it e.g.},
\citealt{jackson_observational_2009}, \citealt{teitler_why_2014},
\citealt{penev_empirical_2018}, \citealt{cameron_hierarchical_2018},
\citealt{hamer_schlaufman_2019}).

An empirical resolution might be possible through long-term timing of
transits and occultations, seeking evidence for changes in the orbital
period.  For instance, long-term transit timing and radial velocity
measurements for WASP-12b have revealed a secular decrease in the
period at a rate of $\approx$30 milliseconds per year, which has been
interpreted as the effect of tidal orbital decay
\citep{maciejewski_departure_2016,patra_2017,maciejewski_planet-star_2018,yee_orbit_2020}.

This study draws attention to a confounding factor that, while
elementary, does not seem to have received the attention it deserves.
The point is that observational programs aimed at identifying orbital
decay in hot Jupiters through transit timing must be accompanied by
concurrent long-term radial velocity monitoring.  The reason is that
an apparent change in period can be produced by the Doppler effect
associated with acceleration of the hot-Jupiter host star along the
line of sight, such as the acceleration that might be produced by a
massive, wide-orbiting companion.  Massive outer companions to hot
Jupiters are common.  \citet{bryan_statistics_2016} calculated an
occurrence rate of $70\pm8\%$ for outer companions to hot Jupiters
with masses from 1-13$\,$$M_{\rm Jup}$ and semi-major axes from
1-20$\,$AU.  Therefore, we expect that many hot Jupiters will display
secular trends in orbital period that are unrelated to tidal orbital
decay.  This possibility can be checked by performing long-term
radial-velocity monitoring at a level sensitive enough to detect or
rule out the relevant amplitude of acceleration.

The focus of this study is the hot Jupiter WASP-4b, which has an
orbital period that appears to be decreasing by about 10 milliseconds
per year.  The period decrease was identified by \citep[][hereafter
\citetalias{bouma_wasp4b_2019}]{bouma_wasp4b_2019}, who combined data
from the NASA TESS mission \citep{ricker_transiting_2015} and a decade
of ground-based transit observations. Soon thereafter,
\citet{southworth_transit_2019} reported an additional 22 transit
times and recalculated the period derivative to be $\dot{P} = -9.2 \pm
1.1$ milliseconds per year.  A separate study by \citet{baluev_2019}
reported on additional transit times, and pointed out that the period
decrease was statistically significant only when analyzing the data
with the highest precision.

To determine the origin of the period change, we acquired four
additional radial velocity measurements with the Keck I 10m telescope
and the High Resolution Echelle Spectrometer (HIRES;
\citealt{vogt_hires_1994}).  In doing so, we extended the time
baseline of HIRES measurements from 3 to 9 years.  The previously
available HIRES data led to the marginal ($\approx$$2\sigma$)
detection of a radial-velocity trend \citep{knutson_friends_2014}.
Our new measurements reveal a line-of-sight acceleration of
$\dot{\gamma} = -0.0422^{+0.0028}_{-0.0027}\ {\rm
m\,s^{-1}\,day^{-1}}$.  Through the Doppler effect\footnote{While the
apparent period change caused by a line-of-sight acceleration has been
referred to as the ``R{\o}mer effect'' \citep{yee_orbit_2020}, a
simpler and probably better term is the Doppler effect. The R{\o}mer
effect refers to the {\it delay} in the reception of a signal due to a
change in the time required for light to traverse the distance between
the source and observer. The Doppler effect refers to the change in
the apparent {\it rate} of a process due to changes in the relative
motion of the source and observer, such as the rate of transits.},
this translates into an expected period decrease of $-5.9$
milliseconds per year, which is comparable to the period decrease that
was measured from transit timing.  We undertook high-resolution
(speckle) imaging to search for evidence of a companion that could be
responsible for the acceleration of WASP-4.

Section~\ref{sec:observations} of this paper presents all of the
available transit data as well as the new radial velocity and speckle
imaging observations.  Section~\ref{sec:analysis} describes our
analysis of the data, and our interpretation that WASP-4 is being
pulled around by a brown dwarf or low-mass star.
Section~\ref{sec:discussion} places this result within the context of
orbital decay searches, and points out that line-of-sight
accelerations will be a relatively common type of ``false positive.''
Section~\ref{sec:conclusions} offers concluding remarks.

\section{Observations}
\label{sec:observations}

\begin{figure*}[t]
	\begin{center}
		\leavevmode
		\includegraphics[width=0.7\textwidth]{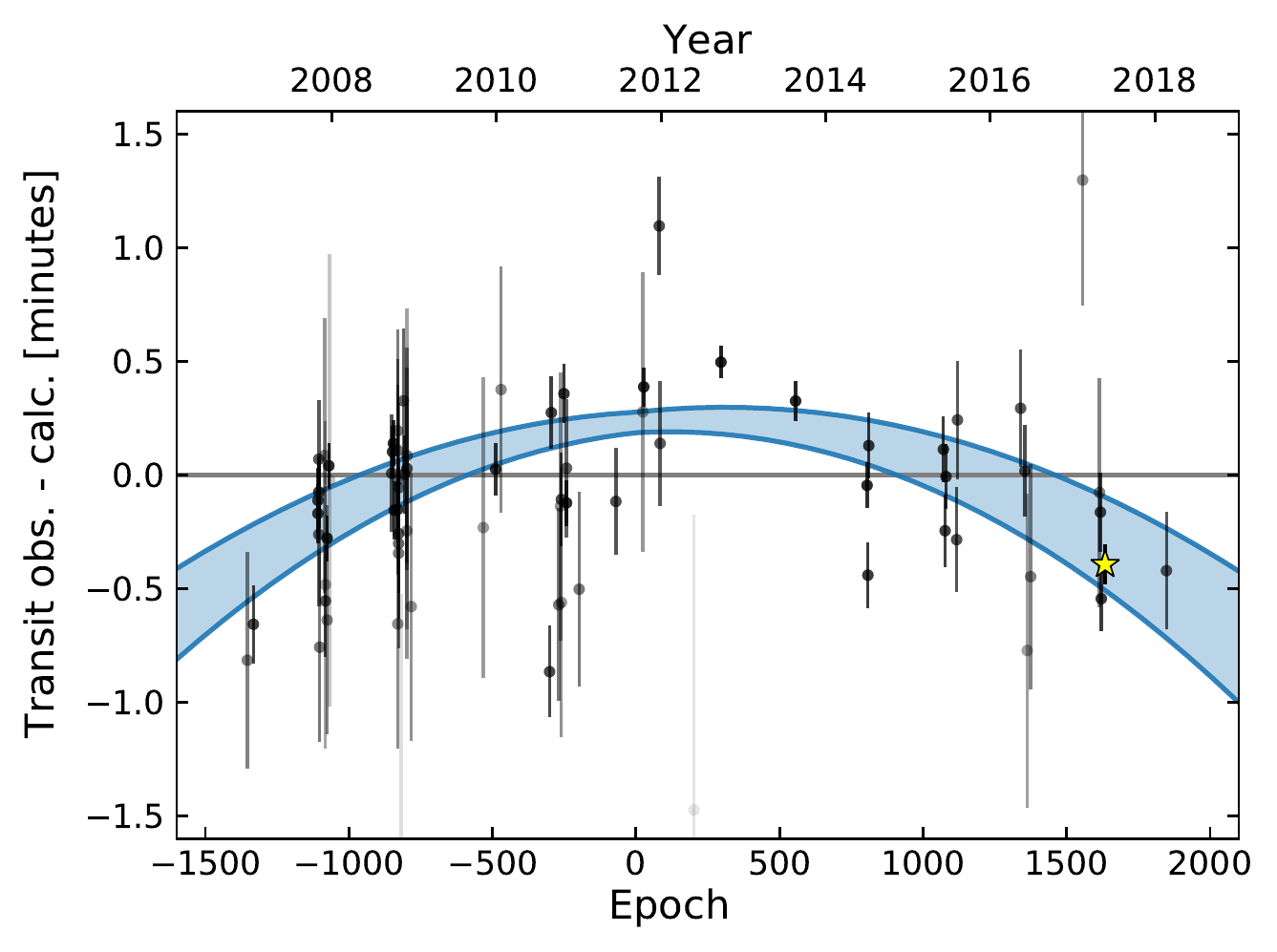}
	\end{center}
	\vspace{-0.7cm}
	\caption{ {\bf Timing residuals and best-fit models for WASP-4b.}
		The vertical axis shows the observed transit times minus the
		calculated times, assuming a constant orbital period.  More opaque
		points correspond to more precise data.  The $\pm1\sigma$
		uncertainties of the quadratic ephemeris are shown in blue.  The
		yellow star represents the weighted average of 18 data points
		obtained with TESS. The TESS data were averaged here for display
		purposes only; our analysis used the 18 individual transit times.
		\label{fig:times}
	}
\end{figure*}

\subsection{Transits}

Table~1 lists the transit times we collected for our analysis.  We
included data from the peer-reviewed literature for which {\it (i)}
the reported time was based on the data from a single transit (as
opposed to fitting the data spanning multiple transits and assuming
the period to be constant), {\it (ii)} the central transit time was
allowed to be a free parameter, and {\it (iii)} the time system was
documented clearly, in particular specifying whether barycentric or
heliocentric corrections had been performed and whether leap seconds
had been taken into account (TDB or UTC).

Most of these data are identical to the data presented by
\citetalias{bouma_wasp4b_2019}.  Twenty-two new times reported by
\citet{southworth_transit_2019} are included.  These transits were
observed from the 3.58m New Technology Telescope and the Danish 1.54m
telescope at La Silla, Chile, and the South African Astronomical
Observatory 1.0m telescope.  Additional timing measurements were also
reported recently by \citet{baluev_2019}, based on a homogeneous
analysis of archival ground-based observations.  We included
\deleted{twelve of} their \deleted{``high quality''} transit times
from the TRAPPIST telescope (six transits), the El~Sauce 36$\,$cm
(four transits), and \citet{petrucci_no_2013} \added{(two transits)}.
For TRAPPIST and El~Sauce, we verified with the original observers
that the timestamps were amenable to the appropriate barycentric and
leap second corrections (M.~Gillon, P.~Evans, priv{.} comm{.}).  We
omitted the fourteen remaining
ETD\footnote{\url{http://var2.astro.cz/ETD}} times from
\citeauthor{baluev_2019} due to ambiguity in whether leap-second
corrections had been performed.  We did not include in our analysis
the four occultation times tabulated by
\citetalias{bouma_wasp4b_2019}, because of the large timing
uncertainties and negligible statistical power.

\subsection{Radial velocities}

We acquired four new radial velocity measurements with Keck/HIRES.
Our observations were performed using the standard setup and reduction
techniques of the California Planet Survey \citep{howard_cps_2010}.
Previously, the HIRES data-points spanned 2010 to 2013
\citep{knutson_friends_2014}.  Our new measurements triple the HIRES
observing baseline to nine years.

The complete set of radial velocity observations is given in Table~2.
Along with the 2010--2019 HIRES observations are early measurements
with two different spectrographs.  \citet{wilson_wasp-4b_2008} and
\citet{triaud_spin-orbit_2010} observed WASP-4 with the Swiss 1.2m
Euler Telescope and CORALIE Spectrograph; we adopted the radial
velocity values from the homogeneous analysis of the latter authors.
We also included data from the High Accuracy Radial Velocity Planet
Searcher (HARPS), reported by \citet{pont_determining_2011} and
\citet{husnoo_observational_2012}.  While
\citet{triaud_spin-orbit_2010} also acquired HARPS data over three
nights for Rossiter-McLaughlin observations, these data were reduced
with a non-standard pipeline making them ill-suited for our study, and
we did not include them.\footnote{This problem was fixed in principle
by \citet{trifonov_public_2020} who performed a homogeneous
re-reduction of the entire HARPS data archive.  We found that the
decision regarding whether to include or omit these points did not
noticeably affect our results.}

\subsection{Speckle imaging}

Once we saw that the new HIRES observations implied a highly
significant trend in the radial velocity, we sought independent
evidence for a wide-orbiting companion by performing speckle imaging
with the Zorro instrument on the Gemini South 8m
telescope~\citep[see][and the instrument
web-pages\footnote{\url{www.gemini.edu/sciops/instruments/alopeke-zorro/}}]{scott_nessi_2018}.
Zorro is a dual-channel speckle interferometer employing narrow-band
filters centered at 562$\,$nm and 832$\,$nm.  

We observed WASP-4 twice, on the night of September 11-12 with
relatively poor seeing (1.2$''$) and also on the night of September
28-29\added{, 2019}.  On each night, we acquired three sets of
$1000\times 60$$\,$msec exposures.  If a companion is present, the
autocorrelation functions of these speckle images would reveal a
characteristic interference pattern. This pattern is then used to
determine the properties of the detected companion and produce a
reconstructed image.  Using the reconstructed speckle images, contrast
curves are produced to determine the 5-$\sigma$ detection limits (see
\citealt{howell_speckle_2011}).  No companions were detected. The data
from the second night, which had better seeing (0.6$''$), led to the
most constraining limit.  The 832$\,$nm limits were the most useful,
given that any faint companion would likely be redder than the host
star.  Therefore, we opted to use the 832$\,$nm September 28-29
contrast limits in the analysis described below.

\section{Analysis}
\label{sec:analysis}

\begin{figure*}[t]
	\begin{center}
		\leavevmode
		\includegraphics[width=0.7\textwidth]{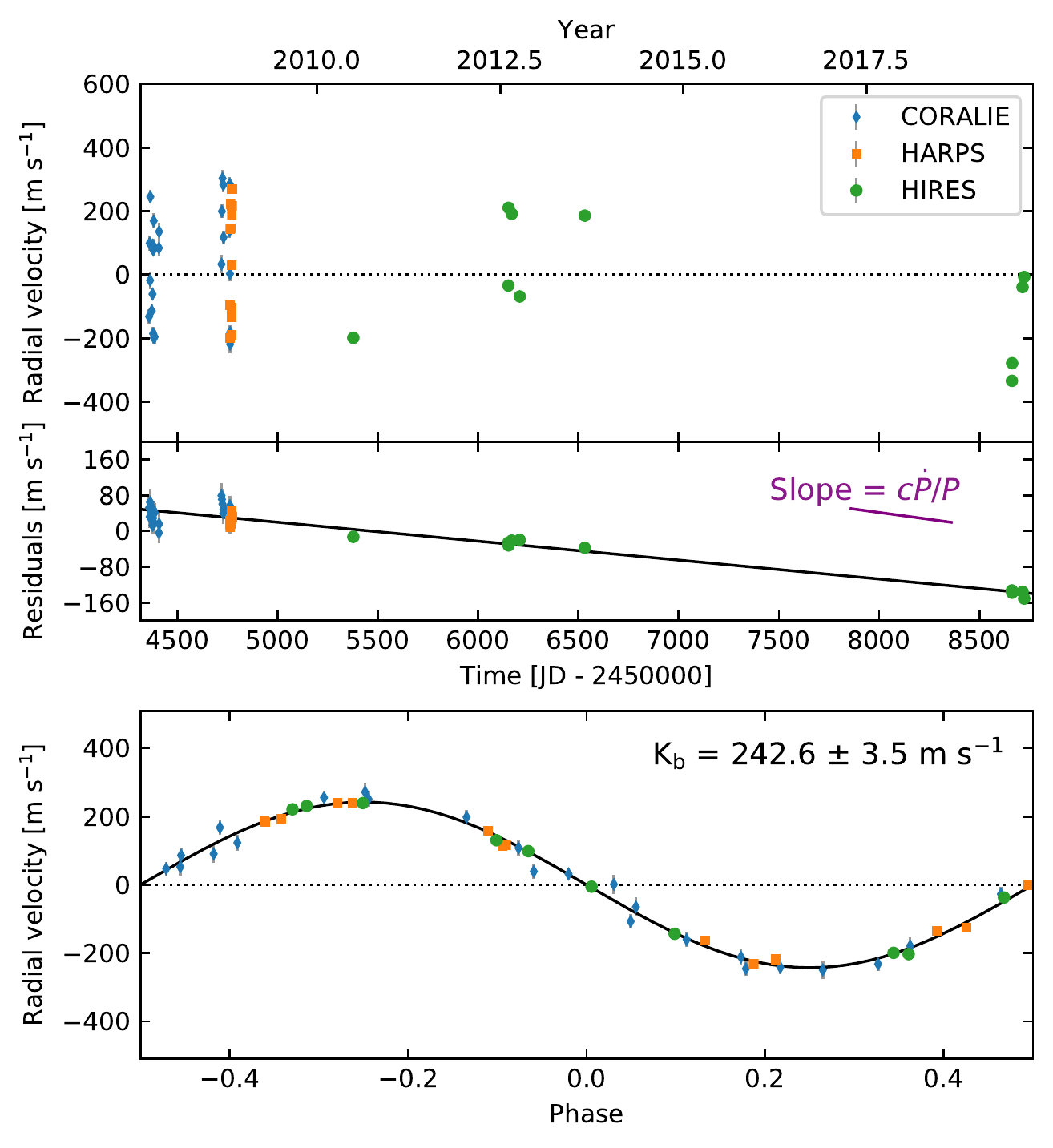}
	\end{center}
	\vspace{-0.7cm}
	\caption{
		{\bf Radial velocities of WASP-4.}
		{\it Top:} RV measurements, with best-fit instrument offsets
		added. 
		{\it Middle:}
		Residuals, after subtracting the best-fitting model for the
		variations induced by the planet WASP-4b. The black line is the
		linear trend inferred from the RV data.  The purple line shows the
		slope that would be needed for the Doppler effect to explain the
		entire period decrease determined from transit timing.  The four
		new RV measurements from this work increase the significance of
		the linear trend from $\approx$2-$\sigma$ to 15-$\sigma$.
		{\it Bottom:}
		Phased orbit of WASP-4b.
		\label{fig:rvs}
		\vspace{-0.0cm}
	}
\end{figure*}

\begin{figure}[!t]
	\begin{center}
		\leavevmode
		\includegraphics[width=0.47\textwidth]{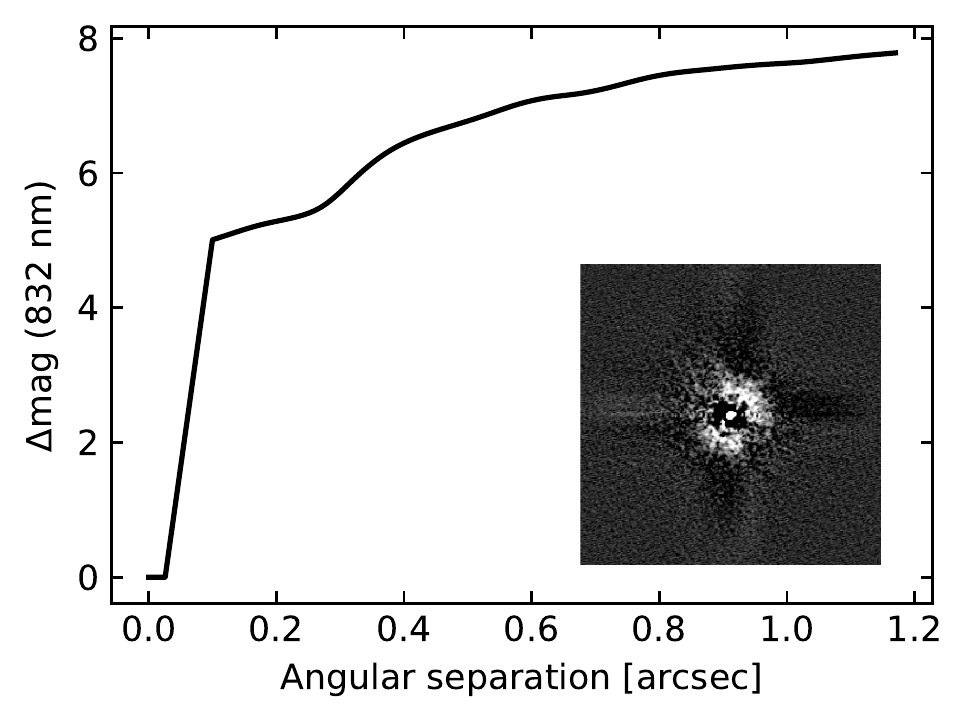}
	\end{center}
	\vspace{-0.7cm}
	\caption{
		{\bf Zorro contrast limits derived from point-source
			injection-recovery experiments.} Sources below the curve would
		have been detected.  The inset shows the speckle image
		reconstructed from 1000 60 millisecond frames in an 832$\,$nm
		bandpass, and acquired on September 28, 2019.  The image scale
		is $2.46''\times2.46''$.
	}
	\label{fig:zorro}
\end{figure}

\begin{figure*}[t]
	\begin{center}
		\leavevmode
		\includegraphics[width=0.8\textwidth]{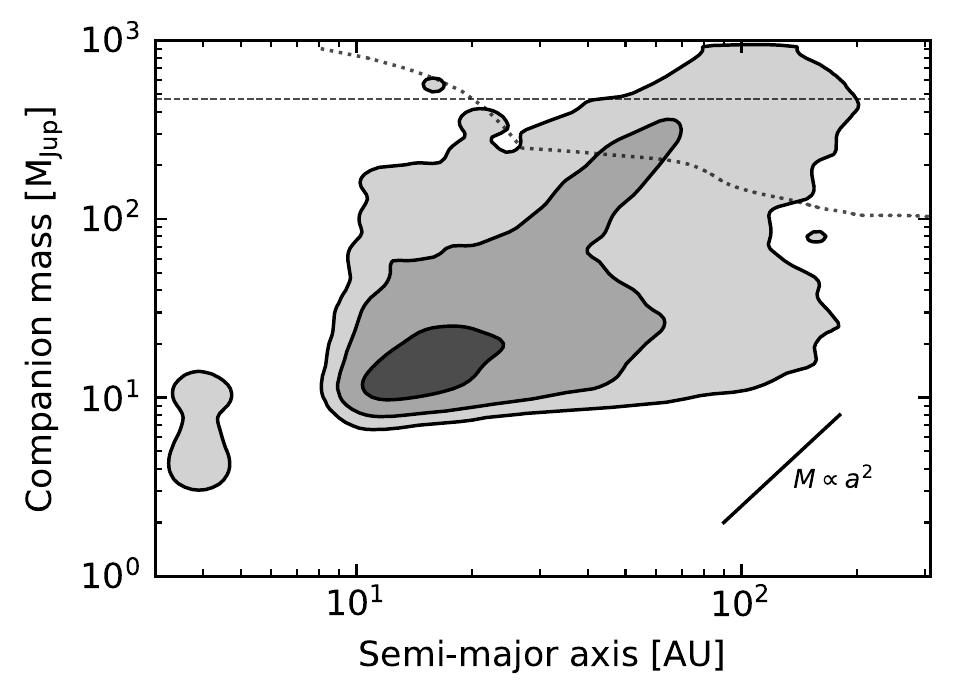}
	\end{center}
	\vspace{-0.8cm}
	\caption{
    {\bf Masses and semi-major axes of companions that meet
    requirements of both the radial velocities and the speckle
    imaging.} Contours show the joint radial velocity and speckle
    imaging probability density with 1, 2, and 3-$\sigma$
    significance. The dotted line shows the speckle imaging limits at
    maximal projected separation.  Relevant projection effects were
    marginalized out when calculating the contours.  The horizontal
    dashed line is the mass limit inferred from observations that
    WASP-4 is single-lined, requiring any companion to contribute no
    more than one-tenth of the observed light.  The black line shows
    the expected degeneracy between mass and semimajor-axis.
    \explain{Caption was modified to account for updated figure. The
    updated upper limit on the mass is now roughly 300$M_{\rm Jup}$,
    instead of 200$M_{\rm Jup}$. The text has been modified
    accordingly.} \label{fig:mass_sma}
		\vspace{-0cm}
	}
\end{figure*}

\subsection{Transits}
\label{sec:transit_analysis}

We fitted two simple timing models to the transit-timing data. The
first model assumes the period $P$ to be constant:
\begin{align}
  t_{\rm tra}(E) &= t_0 + PE,
\end{align}
where $E$ is the integer specifying the number within the sequence of
orbits spanned by the data, and $t_0$ is the transit time for the
event designated $E=0$.  The second model assumes that the period
changes at a steady rate:
\begin{align}
  t_{\rm tra}(E) &=
    t_0 + PE +
    \frac{1}{2} \frac{{\rm d}P}{{\rm d}E} E^2.
  \label{eq:quadratic_model}
\end{align}
The free parameters are the reference time $t_0$, the period at the
reference time $P$, and the period derivative, ${\rm d}P/{\rm d}t =
(1/P) {\rm d}P/{\rm d}E$.  We defined the epoch numbers such that
$E=0$ is near the weighted average of the times of observed transits.
This choice leads to a small covariance between the uncertainties in
$t_0$ and $P$.

We fitted each model by assuming a Gaussian likelihood and sampling
over the posterior probability distributions.  
\added{The timing measurements, uncertainties, and provenances
are given in Table~1.}
We sampled the posterior using the algorithm proposed by
\citet{goodman_ensemble_2010} and implemented by
\citet{foreman-mackey_emcee_2013} in \texttt{emcee}.  The prior for
the quadratic model allowed the period derivative to be either
positive or negative.

Figure~\ref{fig:times} shows the observed transit times, minus the
best-fitting constant-period model.  The best-fitting constant-period
model has 91 degrees of freedom, $\chi^2 = 276$, and $\chi^2_{\rm
red}=3.0$.  The best-fitting quadratic model has 90 degrees of
freedom, $\chi^2 = 183$, and $\chi^2_{\rm red}=2.0$.  The difference
in the Bayesian information criteria (BIC) between the linear and
quadratic and models is $\Delta {\rm BIC} = 89$, strongly favoring the
quadratic model \citep{kass_bayes_1995}.

From the high values of $\chi^2_{\rm red}$, we can surmise that
neither model provides a satisfactory fit to the transit data; there
must be some additional signal or noise.  The earlier study by
\citetalias{bouma_wasp4b_2019} found that the quadratic model for the
(sparser) transit data gave $\chi^2_{\rm red}=1.0$.  The worsening of
$\chi^2_{\rm red}$ \replaced{probably reflects}{could reflect}
underestimated \replaced{measurement}{statistical} uncertainties in at
least some of the newly reported transit times\deleted{, due to
time correlated noise in the photometry}. It might also be due to
\replaced{errors or}{systematic} misunderstandings of the time systems
in which the data were recorded, despite our best efforts to interpret
the literature. \added{In particular, mistaken leap-second corrections
can introduce systematic errors of order one minute.}  \replaced{In
what follows,}{Since we have no way of identifying which observations
are affected by these issues,} we opted to \added{uniformly} enlarge
the uncertainties in the best-fitting parameters of each model by a
factor of $(\chi^2_{\rm red})^{1/2}$.  This was a factor of
$\approx$1.73$\times$ for the linear model, and $\approx$1.41$\times$
for the quadratic model.

The resulting period derivative for the quadratic model is 
\begin{equation}
\dot{P}
  = - (2.74 \pm 0.28)\times 10^{-10}
  = - 8.64 \pm 1.26~{\rm msec}\,{\rm yr}^{-1}.
  \label{eq:dP_dt_obs}
\end{equation}
This agrees to within 1-$\sigma$ of the value reported by
\citet{southworth_transit_2019} ($\dot{P} = - 9.2 \pm 1.1~{\rm
ms}\,{\rm yr}^{-1}$).  It is 2.3-$\sigma$ larger than the rate of
period decrease reported by \citetalias{bouma_wasp4b_2019} ($- 12.6
\pm 1.2~{\rm msec}\,{\rm yr}^{-1}$), presumably because of the new
data from \citeauthor{southworth_transit_2019} and
\citeauthor{baluev_2019}. The other best-fitting model parameters are
reported in Table~3.

\subsection{Radial velocities}
\label{subsec:rvs}

Our initial model for the radial velocity data was a circular orbit
plus instrument offsets, ``jitter'' values (explained below), and a
long-term linear trend \citep[][\texttt{radvel}]{fulton_radvel_2018}.
We set Gaussian priors on the orbital period and time of inferior
conjunction using the values from Table~4 of
\citetalias{bouma_wasp4b_2019}.  We assumed the orbit to be circular
based on previous studies that placed stringent upper limits on the
eccentricity
\citep{beerer_secondary_2011,knutson_friends_2014,bonomo_gaps_2017}.
The free parameters were the orbital velocity semi-amplitude, the
instrument zero-points, a ``white noise'' instrument jitter for each
instrument added in quadrature to its uncertainties, and a linear
($\dot{v_{\rm r}}$) acceleration term.  (Without the linear trend, the
model is a much poorer fit with $\Delta {\rm BIC}=73$.)

In the best-fitting model, WASP-4 is accelerating along our
line-of-sight at a rate
\begin{equation}
  \dot{v}_{\rm r} = \dot{\gamma} = 
     -0.0422^{+0.0028}_{-0.0027}
     \,{\rm m}\,{\rm s}^{-1}\,{\rm day}^{-1}.
\end{equation}
The other model parameters are listed in Table~4.  Based on earlier
data, $\dot{\gamma}$ was thought to be about five times smaller, and
had marginal statistical significance
\citep{knutson_friends_2014,bouma_wasp4b_2019}.

Because of the Doppler effect, any line-of-sight acceleration should
lead to a change in the observed orbital period:
\begin{align}
  \dot{P}_{\rm\,RV} &= \frac{\dot{v}_{\rm r} P}{c},
\end{align}
or in more convenient units,
\begin{align}
  \dot{P}_{\rm\,RV} &= 105{.}3\,{\rm msec}\,{\rm yr}^{-1} \
  \left( \frac{P}{{\rm day}} \right)
  \left( \frac{\dot{\gamma}}{ {\rm m}\,{\rm s}^{-1}\,{\rm day}^{-1}} \right).
\end{align}
For WASP-4, this yields
\begin{align}
  \dot{P}_{\rm\,RV} &= -5.94 \pm 0.39\,{\rm msec}\,{\rm yr}^{-1}.
\end{align}
Therefore, most of the period derivative that was detected through
transit timing ($\dot{P}= - 8.64 \pm 1.26~{\rm msec}\,{\rm yr}^{-1}$)
can be accounted for by the Doppler effect.  Given the evidence for
unmodeled noise in both the transit timing data and the
radial-velocity data, it seems plausible that the Doppler effect can
account for the entire observed period derivative.

An important consideration is whether the measured RV trend is truly
due to acceleration or whether it is due to stellar activity.  We
investigate this by analyzing the Ca II H \& K lines in the WASP-4
spectra, as quantified with the chromospheric $S$-index
\citep{wright_chromospheric_2004}.  We relied only on the HIRES
velocities, which were the crucial source of information in the
radial-velocity analysis.  First, we subtracted the component of the
best-fitting model representing the orbital motion induced by the
planet.  Then, following
\citet{bryan_statistics_2016,bryan_excess_2019}, we calculated the
Spearman rank correlation coefficient between the $S$-index and the
orbit-subtracted velocities.  We found a correlation coefficient of
0.16. This correlation is not statistically significant; the
corresponding $p$-value is 0.65.  Furthermore, inspection of the
$S$-index timeseries did not show secular or sinusoidal trends, as
would be expected if we were observing a long-term magnetic activity
cycle.  The $S$-index values are included in Table~2.  We conclude
that there is no evidence that the linear trend is caused by stellar
activity.

\subsection{Constraints on companion masses and semi-major axes}

Given a linear radial velocity trend, we can place probabilistic
constraints on the mass and semi-major axis of the additional body
that is causing the trend.  For a quick estimate of the minimum mass
required to explain the linear trend in WASP-4, we turned to
\citet{feng_california_2015}.  As they discussed, the scenario that
yields the minimum companion mass for a system with a linear trend is
a companion with $e\approx0.5$ and $\omega=90^\circ$.  Substituting
$P\approx 1.25\tau$ and $K \approx 0.5\tau \dot{\gamma}$ into the mass
function \citep[{\it e.g.},][]{wright_efficient_2009} yields
\begin{equation}
 M_{\rm min} \approx 5.99\,M_{\rm Jup}
  \left( \frac{\tau}{{\rm yr}} \right)^{4/3}
  \left| \frac{\dot{\gamma}}{{\rm m\,s^{-1}\,day^{-1}}} \right|
  \left( \frac{M_\star}{{M_\odot}} \right)^{2/3},
\end{equation}
where $\tau$ is the observing baseline.  For WASP-4, this gives
$M_{\rm min} = 4.9\,M_{\rm Jup}$.  Higher masses are allowed for
companions that orbit further from the star: at fixed $\dot{\gamma}$,
$M_{\rm comp} \propto a^2$
\citep{torres_substellar_1999,liu_crossing_2002}.

High-resolution images can further limit the available parameter space
by setting an upper limit on the companion brightness (and the
corresponding mass) as a function of orbital distance.  The procedure
we used to combine constraints from both radial velocities and high
resolution imaging was developed by \citet{wright_linear_trends_2007},
\citet{crepp_trends_2012}, \citet{montet_trends_2014},
\citet{knutson_friends_2014},
\citet{bryan_statistics_2016,bryan_excess_2019}, and others.

\paragraph{Speckle imaging transformations}

First, we converted the contrast curves obtained from speckle imaging
(Figure~\ref{fig:zorro}) to upper limits on the companion mass as
a function of \added{projected} separation. To do this, we followed
\citet{montet_trends_2014}, and opted to employ the
\citet{baraffe_evolutionary_2003} models for substellar mass objects
and the MIST isochrones for stellar mass objects
\citep{paxton_modules_2011,paxton_modules_2013,paxton_modules_2015,dotter_mesa_2016,choi_mesa_2016}.
We assumed that the system age was 5 Gyr, at which point the companion
would have fully contracted.

Due to the custom filters of the Zorro imager, and corresponding lack
of synthetic photometry, we further assumed that all sources had
blackbody spectra. While this is a simplification, we do not have
ready access to the planetary and stellar atmosphere models needed for
the consistent calculation with the \texttt{COND03} and \texttt{MESA}
models.  We adopted the effective temperatures and bolometric
luminosities from the \citet{baraffe_evolutionary_2003} and MIST
isochrones.  Using these theoretical quantities and the
empirically-measured Zorro bandpasses, we calculated absolute
magnitudes in the 562 and 832 nm Zorro bands for stellar and planetary
mass companions.  Applying the same calculation to WASP-4 itself using
the effective temperature and bolometric luminosity from
\citetalias{bouma_wasp4b_2019}, we derived the transformation from
contrast ratio to companion mass.  The resulting
limits \added{derived if we assume maximal projected separations} are
\replaced{shown}{shown with the dotted line} in Figure~\ref{fig:mass_sma}.
\added{
However, because the primary star is accelerating towards
our line of sight, the companion could very well be near
inferior conjunction.
Our approach for incorporating the relevant projection effects
is described in the following paragraphs.}

\paragraph{Combined radial velocity and imaging constraints}
To derive constraints on possible companion masses and separations
\deleted{from the radial velocities}, we mostly followed the procedure of
\citet{bryan_excess_2019}. We began by defining a $128\times128$ grid
in true planetary mass and semimajor axis, over a logarithmic grid
ranging from 1 to 900$\,$$M_{\rm Jup}$ and 3 to 500$\,$${\rm AU}$.  We
then considered the possibility that an additional companion in any
particular cell could explain the observed linear trend.

For each grid cell, we simulated 512 hypothetical companions. We
assigned each companion a mass and semimajor axis from log-uniform
distributions within the boundaries of the grid cell. We drew the
inclination from a uniform distribution in $\cos i$.  For companion
masses less than $10\,M_{\rm Jup}$, we drew the eccentricity from
\citet{kipping_beta_2013}'s long-period exoplanet Beta distribution
($a=1.12$, $b=3.09$).  If the companion mass exceeded $10\,M_{\rm
Jup}$, we drew the eccentricity from the power-law $p_e \propto
e^\eta$ reported by \citet{moe_mind_2017} in their Equation~17 ($\eta
\approx 0.5$ for most orbital periods).  The long-period exoplanet and
long-period binary eccentricity distributions are quite different: the
exoplanet distribution is ``bottom-heavy'', with eccentricities
preferentially close to zero.  The binary star distribution is
``top-heavy'', with a broad range of eccentricities extending close to
unity \citep{moe_mind_2017,price-whelan_close_2020}.  The choice of
$10\,M_{\rm Jup}$ as the dividing line between these two regimes was
based on the empirical study of \citet{schlaufman_evidence_2018} on
the distinction between giant planets and brown dwarfs. This value is
also close to the $13\,M_{\rm Jup}$ deuterium-burning limit
\citep[{\it e.g.},][]{burrows_nongray_1997}.

\explain{Modified wording for clarity.}
\replaced{For each simulated companion, we drew a set of model
parameters from the converged posterior probability samples of our
model for the WASP-4 radial velocities.}{The orbital properties of the
inner hot Jupiter were assigned for each simulated system by sampling
from from the radial velocity posterior derived in
Section~\ref{subsec:rvs}.}  We subtracted the orbital component of the
model from the observed RVs, leaving behind the RV residuals with a
linear trend.  Given $(a_{\rm c}, M_{\rm c}, e_{\rm c})$ for each
simulated outer companion, and the choice of instrument offsets and
jitters, we performed a maximum likelihood fit for the time and
argument of periastron of the outer simulated companion.  \deleted{We
converted the resulting $128\times128\times512$ cube of log-likelihood
values to probabilities, and averaged over the samples in each grid
cell to derive a probability distribution in mass and semi-major axis,
based on the radial velocities.}

\added{We then incorporated the speckle imaging limits in each
simulated system as follows. After fitting for the time and argument
of periastron, all the orbital parameters needed to find the projected
separation at the time of observation are known. We assumed uniform
sensitivity as a function of position angle, and therefore fixed the
longitude of the ascending node to zero. We then calculated the
projected separation using the parametrization given by
\citet{quirrenbach_2010}.  If a simulated companion's mass and
projected separation put it above the 5$\sigma$ contrast curve, we
assumed it would have been detected.  We multiplied the resulting
$128\times128\times512$ cubes of radial velocity and speckle imaging
probabilities, and marginalized over the systems in each grid cell to
derive a probability distribution in mass and semi-major axis.
}
\replaced{Figure~\ref{fig:mass_sma} shows the result.}
{The contours in Figure~\ref{fig:mass_sma} show the result: the companion
responsible for the acceleration has a true mass of 10-300$\,M_{\rm
Jup}$ and an orbital distance of $10$-$100\,{\rm AU}$.}

\section{Discussion}
\label{sec:discussion}

\subsection{Implications for WASP-4}

The previously offered explanations for WASP-4b's decreasing orbital
period included tidal orbital decay, apsidal precession, and the
Doppler effect \citep{bouma_wasp4b_2019}.  Our new radial velocity
measurements strongly indicate that the least exotic option---the
Doppler effect---is the \replaced{most likely explanation}{dominant
physical process}.  Long-term transit timing data show the orbital
period to be decreasing by $- 8.64 \pm 1.26~{\rm msec}\,{\rm
yr}^{-1}$.  The long-term radial velocity data show a trend that
should lead to an apparent period derivative of $-5.94 \pm 0.39~{\rm
msec}\,{\rm yr}^{-1}$ through the Doppler effect.  Although these two
measurements of the period derivative are discrepant by about
2-$\sigma$, Occam's razor \replaced{suggests that most or all
of}{would suggest that} the apparent decrease of WASP-4b's orbital
period is caused \added{solely} by the line-of-sight
acceleration.\added{ Detection of additional second-order effects will
require more precise data, and a more significant discrepancy.}

Based on the data, the companion causing the acceleration is probably
either a brown-dwarf or low mass star with an orbital distance of
$10$-$100\,{\rm AU}$ (Figure~\ref{fig:mass_sma}).  Given such a mass,
this companion may have influenced the orbital evolution of the hot
Jupiter orbiting WASP-4, as well as any other planets in the system.
The fact that most hot Jupiters have similar massive outer companions
\citep{knutson_friends_2014,bryan_statistics_2016} is compatible with
some of the high-eccentricity formation theories for hot Jupiters (see
\citealt{dawson_johnson_2018}).  Further radial velocity monitoring
should eventually reveal the orbital parameters and minimum mass of
WASP-4's massive outer companion.

Based on a priori expectations, if the outer companion lives within
100$\,$AU, then it seems more likely to be a brown-dwarf than a
low-mass star.  The reason is that stellar companions within 100$\,$AU
seem to be rare in systems with transiting planets, relative to field
stars.  This issue has been reviewed by \citet{moe_impact_2019}, who
synthesized work by \citet{wang_influence_2014,wang_influence_2015},
\citet{ngo_2015_fohj2}, \citet{ngo_friends_2016},
\citet{kraus_impact_2016}, \citet{matson_stellar_2018},
\citet{ziegler_soar_2020}, and others.  A handful of systems with
inner hot Jupiters and outer brown dwarfs within 100$\,$AU are known,
for instance CoRoT-20, HATS-59, WASP-53, and WASP-81
\citep{triaud_peculiar_2017,rey_brown_2018,sarkis_hats-59bc_2018}.  In
contrast, from surveys by \citet{knutson_friends_2014},
\citet{ngo_2015_fohj2} and \citet{mugrauer_search_2019} we could find
only one example of a system with an inner hot Jupiter and an outer
low-mass star within 100$\,$AU: HAT-P-10 \citep{bakos_2009_hatp10}.
Overall, this body of literature indicates that the presence of a
stellar-mass companion within 100$\,$AU of the host star could hinder
the formation of planetary systems.  If true, then one would expect
that the outer companion in WASP-4 would have the lowest mass allowed
by the data.

\subsection{How many other hot Jupiters are accelerating towards us?}

\begin{figure}[t]
	\begin{center}
		\leavevmode
		\includegraphics[width=0.48\textwidth]{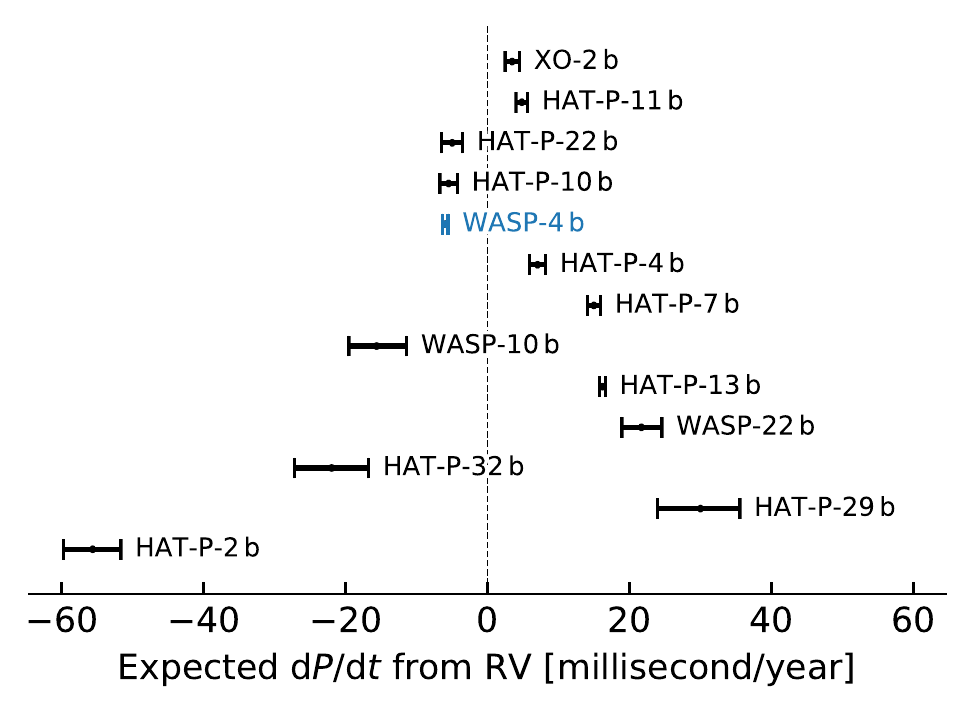}
	\end{center}
	\vspace{-0.4cm}
	\caption{
  {\bf Predicted hot Jupiter period changes from linear radial
  velocity trends.} Including WASP-4b, 16 of 51 hot Jupiters from
  \citet{knutson_friends_2014} have shown long-term radial velocity
  trends.  HAT-P-11 is shown, though its signal is somewhat correlated
  with stellar activity \citep{yee_hat-p-11_2018}.  Three hot Jupiters
  are not shown because their radial velocity curves are better
  described as quadratic trends in time: HAT-P-17, WASP-8, and
  WASP-34.  Objects are ordered in the $y$ dimension by the absolute
  value of d$P$/d$t$.
	\label{fig:pdot_pop}
  \vspace{-0.3cm}
	}
\end{figure}

We identified WASP-4b's decreasing orbital period as part of a search
for tidal orbital decay.  However, most hot Jupiters have companions
outside of $5\,{\rm AU}$ with super-Jovian masses
\citep{knutson_friends_2014,bryan_statistics_2016}.  Line-of-sight
accelerations (both positive and negative) should therefore be common
in hot Jupiter systems. 

To evaluate the importance of these effects for future transit timing
analyses, we collected the linear radial velocity trends reported by
\citet{knutson_friends_2014}, and computed the expected orbital period
derivatives $\dot{P}_{\rm\,RV} = \dot{v}_{\rm r} P/c$ for each system.
The results are given in Table~5, and visualized for hot Jupiters with
significant ($>$$3\sigma$) linear trends in Figure~\ref{fig:pdot_pop}.

Including WASP-4b, 16 of 51 hot Jupiters surveyed by
\citet{knutson_friends_2014} show a non-zero radial velocity trend.
Therefore, around 1 in 3 hot Jupiters \replaced{is}{are} expected to
show period changes comparable to that of WASP-4 due to acceleration
by outer companions.  The sign of the accelerations should be random,
with about half of them approaching and half of them receding.  With a
large enough sample of short-period hot Jupiters, one might be able to
distinguish line-of-sight accelerations from tidal orbital decay by
seeking evidence for a systematic tendency for the observed period to
decrease, rather than increase.

\subsection{At what rate is the measurement precision of d$P$/d$t$
increasing?}
\label{sec:fisher}

For hot Jupiters that have been monitored over baselines exceeding 10
years, secular changes in their orbital periods are currently being
constrained to a precision of $\lesssim$$10~{\rm msec}\,{\rm yr}^{-1}$
\citep{wilkins_searching_2017,maciejewski_planet-star_2018,baluev_2019,petrucci_discarding_2020,patra_2020}.
This is roughly commensurate with the level of signal many outer
companions are expected to induce (Figure~\ref{fig:pdot_pop}).

At what point in time will further detections of the Doppler effect
become routine for hot Jupiters?  More specifically, at what rate does
the uncertainty in the quadratic term of
Equation~\ref{eq:quadratic_model} scale with the observing baseline?
This can be answered with a Fisher analysis of the model
\begin{align}
  t_{\rm tra} = a_0 + a_1 E + a_2 E^2,
\end{align}
where $a_0\equiv t_0$, $a_1\equiv P$, and $a_2\equiv 0.5\cdot {\rm
d}P/{\rm d}E$.  Following \citet{gould_chi2_2003}, one can show that
if $N$ transit timing measurements are taken uniformly across a
baseline of $\Delta E$ epochs with constant precision $\sigma$, then
the uncertainty of the quadratic term is given by
\begin{align}
  \sigma_{a_2} = 6\sqrt{5}
   \frac{\sigma}{N^{1/2} (\Delta E)^2} \propto (\Delta E)^{-5/2}.
\end{align}
This result implies that a doubled observing baseline yields an
$\approx$5.7-fold improvement in precision on d$P$/d$t$.  If regular
observations continue from ground and space-based observatories,
period derivatives will be measured with precision exceeding $1~{\rm
msec}\,{\rm yr}^{-1}$ within the coming decade.

\section{Conclusions}
\label{sec:conclusions}

From newly acquired radial velocity measurements, we found that WASP-4
is accelerating towards the Earth at $\dot{\gamma} =
-0.0422^{+0.0028}_{-0.0027}\ {\rm m}\,{\rm s}^{-1}\,{\rm day}^{-1}$.
The corresponding Doppler effect predicts a period decrease
$\dot{\gamma} P/c$ of $-5.94 \pm 0.39~{\rm msec}\,{\rm yr}^{-1}$.  The
majority of the period decrease observed in transits ($\dot{P}= - 8.64
\pm 1.26~{\rm msec}\,{\rm yr}^{-1}$) is therefore explained by the
acceleration of the host star~---~leaving no evidence for tidal
orbital decay or apsidal precession.  
A probabilistic analysis of the speckle imaging limits and the
radial velocity trend showed that the companion causing the
acceleration is most likely a brown dwarf or low-mass star with
semi-major axis between 10-100$\,$AU.

Most hot Jupiters have outer companions with masses larger than
Jupiter beyond 5$\,$AU
\citep{knutson_friends_2014,bryan_statistics_2016}. The accelerations
and period changes induced by these outer companions will become an
increasingly large nuisance in the hunt for tidal orbital decay as the
observational baselines get longer.  In particular, the precision with
which the period derivative can be measured from transits scales with
the baseline duration to the 5/2 power (Section~\ref{sec:fisher}).
Within a decade, many more hot Jupiters should show orbital period
changes due to accelerations from their outer companions.  To
distinguish this effect from tidal decay, further long-term radial
velocity measurements of hot Jupiters are strongly encouraged.

\software{
  \texttt{astrobase} \citep{bhatti_astrobase_2018},
  \texttt{astropy} \citep{astropy_2018},
  \texttt{astroquery} \citep{astroquery_2018},
  \texttt{corner} \citep{corner_2016},
  \texttt{emcee} \citep{foreman-mackey_emcee_2013},
  \texttt{IPython} \citep{perez_2007},
  \texttt{matplotlib} \citep{hunter_matplotlib_2007}, 
  \texttt{MESA} \citep{paxton_modules_2011,paxton_modules_2013,paxton_modules_2015}
  \texttt{numpy} \citep{walt_numpy_2011}, 
  \texttt{pandas} \citep{mckinney-proc-scipy-2010},
  \texttt{radvel} \citep{fulton_radvel_2018},
  \texttt{scipy} \citep{jones_scipy_2001}.
}

\facilities{
	{\it Astrometry}:
	Gaia \citep{gaia_collaboration_gaia_2016,gaia_collaboration_gaia_2018}.
	{\it Imaging}:
	Gemini:South~(Zorro; \citealt{scott_nessi_2018}.
	{\it Spectroscopy}:
	Keck:I~(HIRES; \citealt{vogt_hires_1994}),
	Euler1.2m~(CORALIE),
	ESO:3.6m~(HARPS; \citealt{mayor_setting_2003}).
	{\it Photometry}:
	CTIO:1.0m (Y4KCam),
	Danish 1.54m Telescope,
	El Sauce:0.356m,
	Elizabeth 1.0m at SAAO,
	Euler1.2m (EulerCam),
	Magellan:Baade (MagIC),
	Max Planck:2.2m	(GROND; \citealt{greiner_grond7-channel_2008})
	NTT,
	SOAR (SOI),
	TESS \citep{ricker_transiting_2015},
	TRAPPIST \citep{jehin_trappist_2011},
	VLT:Antu (FORS2).
}

%
%
\nocite{wilson_wasp-4b_2008}
\nocite{gillon_improved_2009}
\nocite{winn_transit_2009}
\nocite{hoyer_tramos_2013}
\nocite{dragomir_terms_2011}
\nocite{sanchis-ojeda_starspots_2011}
\nocite{nikolov_wasp-4b_2012}
\nocite{ranjan_atmospheric_2014}
\nocite{huitson_gemini_2017}

\input{WASP-4b_transit_time_table.tex}
\input{WASP-4b_rv_table.tex}

\input{model_fit_table.tex}
\input{rv_model_posterior_table.tex}
\input{pdot_table.tex}

\bibliographystyle{yahapj}                            
\bibliography{bibliography}

\listofchanges

\end{document}

%% file: WASP-4b_transit_time_table.tex
\startlongtable
\begin{deluxetable*}{ccccc}
    

\tabletypesize{\scriptsize}


\tablecaption{WASP-4b transit times.}
\label{tab:transit_times}

\tablenum{1}

\tablehead{
  \colhead{$t_{\rm tra}$ [BJD$_\mathrm{TDB}$]} &
  \colhead{$\sigma_{t_{\rm tra}}$ [days]} &
  \colhead{Epoch} & 
  \colhead{Time Reference} & 
  \colhead{Observation Reference}
}

\startdata
 2454368.59279 &      0.00033 &   -1354 &       \citet{hoyer_tramos_2013} &           \citet{wilson_wasp-4b_2008} \\
\enddata


\tablecomments{
Table~1 is published in its entirety in a machine-readable format.
The first row is shown for guidance regarding form and
content.  $t_{\rm tra}$ is the measured transit midtime, and
$\sigma_{t_{\rm tra}}$ is its $1\sigma$ uncertainty.  ``Time
Reference'' refers to the provenance of the timing measurement, which
may differ from the ``Observation Reference'' in cases for which a
homogeneous timing analysis was performed.  The
\citealt{hoyer_tramos_2013} BJD$_{\rm TT}$ times are equal to
BJD$_{\rm TDB}$ for our purposes \citep{urban_explanatory_2012}.
}
\vspace{-1.5cm}
\end{deluxetable*}

%% file: WASP-4b_rv_table.tex
\startlongtable
\begin{deluxetable*}{cccccc}
    

\tabletypesize{\scriptsize}


\tablecaption{WASP-4b radial velocities.}
\label{tab:rvs}

\tablenum{2}

\tablehead{
  \colhead{Time [BJD$_\mathrm{TDB}$]} &
  \colhead{RV [m$\,$s$^{-1}$]} &
  \colhead{$\sigma_{\rm RV}$ [m$\,$s$^{-1}$]} & 
  \colhead{$S$-value} &
  \colhead{Instrument} & 
  \colhead{Provenance}
}

\startdata
 2454321.12345 &      42 &   0.42 &       0.42    & HIRES & \citet{knutson_friends_2014} \\
\enddata


\tablecomments{
Table~2 is published in its entirety in a machine-readable format.
The first entry is shown for guidance regarding form and
content. $S$-values are reported only for the HIRES measurements.
}
\vspace{-0.9cm}
\end{deluxetable*}

%% file: model_fit_table.tex
\startlongtable
\begin{deluxetable*}{lc}

\tabletypesize{\footnotesize}

\tablenum{3}


\tablecaption{Best-fit transit timing model parameters.}
\label{tab:bestfit}

\tablehead{
  \colhead{Parameter} &
  \colhead{Median Value~(Unc.)\tablenotemark{a}}
}
\startdata
~~~~~~{\it Constant period} &  \\
$t_0$\,[${\rm BJD}_{\rm TBD}$]    & 2456180.558712(+24)(-24)              \\
$P$\,[days]                       & 1.338231429(+26)(-26)                 \\
~~~~~~{\it Constant period derivative} &  \\
$t_0$~[${\rm BJD}_{\rm TBD}$]     & 2456180.558872(+31)(-31)              \\
$P$\,[days]                       & 1.338231502(+24)(-24)                 \\
$dP/dt$                           & $-2.74(+40)(-40) \times 10^{-10}$     \\
\enddata
\tablenotetext{a}{
The numbers in parenthesis give the $68\%$ confidence interval for the
final two digits, where appropriate.  The intervals have been inflated
by a factor of $(\chi^2_{\rm red})^{1/2}$ due to excess scatter in the
transit residuals (see Section~\ref{sec:transit_analysis}).
}
\vspace{-2cm}
\end{deluxetable*}

%% file: rv_model_posterior_table.tex
\startlongtable
\begin{deluxetable*}{lrrr}
\tablecaption{ Best-fit radial velocity model parameters. }
\label{tab:params}
\tablenum{4}
\tablehead{
  \colhead{Parameter} & 
  \colhead{Credible Interval} & 
  \colhead{Maximum Likelihood} & 
  \colhead{Units}
}
\startdata
\sidehead{~~~~~\it{Orbital Parameters}}
  $P_{b}$ & $1.338231466\pm 2.3e-08$ & $1.338231466$ & day \\
  $T\rm{conj}_{b}$ & $2455804.515752^{+2.5e-05}_{-2.4e-05}$ & $2455804.515752$ & BJD$_{\rm TDB}$ \\
  $e_{b}$ & $\equiv0.0$ & $\equiv0.0$ &  \\
  $\omega_{b}$ & $\equiv0.0$ & $\equiv0.0$ & $^\circ$ \\
  $K_{b}$ & $242.6^{+3.6}_{-3.5}$ & $242.6$ & m$\,{\rm s}^{-1}$ \\
\sidehead{~~~~~\it{Other Parameters}}
  $\gamma_{\rm HIRES}$ & $36.4^{+5.8}_{-5.9}$ & $36.4$ & m$\,{\rm s}^{-1}$ \\
  $\gamma_{\rm HARPS}$ & $-69.9^{+4.2}_{-4.1}$ & $-70.1$ & m$\,{\rm s}^{-1}$ \\
  $\gamma_{\rm CORALIE}$ & $-39.9^{+5.5}_{-5.2}$ & $-40.1$ & m$\,{\rm s}^{-1}$ \\
  $\dot{\gamma}$ & $-0.0422^{+0.0028}_{-0.0027}$ & $-0.0424$ & m$\,{\rm s}^{-1}\,{\rm day}^{-1}$ \\
  $\ddot{\gamma}$ & $\equiv0.0$ & $\equiv0.0$ &  \\
  $\sigma_{\rm HIRES}$ & $10.8^{+3.7}_{-2.7}$ & $8.2$ & $\rm m\,s^{-1}$ \\
  $\sigma_{\rm HARPS}$ & $13.0^{+3.7}_{-2.6}$ & $11.5$ & $\rm m\,s^{-1}$ \\
  $\sigma_{\rm CORALIE}$ & $13.8^{+6.6}_{-6.7}$ & $12.9$ & $\rm m\,s^{-1}$ \\
\enddata
\tablenotetext{}{
  Reference epoch for $\gamma$,$\dot{\gamma}$,$\ddot{\gamma}$: 2455470 
}
\vspace{-2.5cm}
\end{deluxetable*}

%% file: pdot_table.tex
\startlongtable
\begin{deluxetable*}{lllllllll}
\tablecaption{
	Predicted hot Jupiter period changes from linear radial
	velocity trends reported by \citet{knutson_friends_2014}.
\label{tab:pdot_table}
}
\tablenum{5}
%
\tablehead{
  \colhead{Planet} &
  \colhead{$\dot{\gamma}$ [m$\,$s$^{-1}$$\,$yr$^{-1}$]} &
  \colhead{$+\sigma_{\dot{\gamma}}$ [m$\,$s$^{-1}$yr$^{-1}$]} & 
  \colhead{$-\sigma_{\dot{\gamma}}$ [m$\,$s$^{-1}$yr$^{-1}$]} & 
  \colhead{$P$ [days]} &
  \colhead{$\dot{P}_{\,{\rm RV}}$ [ms$\,$yr$^{-1}$]} &
  \colhead{$+\sigma_{\dot{P}_{\,{\rm RV}}}$ [ms$\,$yr$^{-1}$]} &
  \colhead{$-\sigma_{\dot{P}_{\,{\rm RV}}}$ [ms$\,$yr$^{-1}$]} &
  \colhead{Significant?}
}
%
\startdata
HAT-P-2 b & -0.0938 & 0.0067 & 0.0069 & 5.6335158 & -55.62 & 3.97 & 4.09 & 1
\enddata
\tablecomments{
  Table~5 is published in its entirety in a machine-readable format.
  The first entry is shown for guidance regarding form and content.
  Orbital periods were retrieved from NASA's Exoplanet Archive.
  Additional comments regarding non-linear trends and stellar activity
  are included in the MRT.
}
\vspace{-0.5cm}
\end{deluxetable*}